# ABORDAGEM PROBABILÍSTICA PARA ANÁLISE DE CONFIABILIDADE DE DADOS GERADOS EM SEQUENCIAMENTOS MULTIPLEX NA PLATAFORMA ABI SOLID

**Fabio M. F. Lobato[1], Carlos D. N. Damasceno[1],**
**Péricles L. Machado[1], Nandamudi L. Vijaykumar[2], André R. dos Santos[3], Sylvain H. Darnet[3], André N. A. Gonçalves[3], Dayse O. de Alencar[3], Ádamo L. de Santana[1]**

[1]Laboratório de Planejamento de Redes de Alto Desempenho - LPRAD
[3]Laboratorio de Genética Humana e Medica - LGHM
Universidade Federal do Pará (UFPA)
Caixa Postal 8619 – 66.075-110 – Belém – PA – Brasil

[2]Laboratório Associado de Computação e Matemática Aplicada
Instituto Nacional de Pesquisas Espaciais (INPE)
Caixa Postal 515 – 12.227-010 – São José dos Campos – SP – Brasil

lobato.fabio@ufpa.br, carlos.damasceno@icen.ufpa.br, pericles.machado@itec.ufpa.br, vijay@lac.inpe.br, andremrsantos@gmail.com, sylvain@ufpa.br, anicolau@ufpa.br, doalencar@yahoo.com.br, adamo@ufpa.br

## RESUMO

O sequenciadores de nova geração como as plataformas Illumina e SOLiD geram uma grande quantidade de dados, comumente, acima de 10 Gigabytes de arquivos-texto. Particularmente, a plataforma SOLiD permite o sequenciamento de múltiplas amostras em uma única corrida, denominada de corrida multiplex, por meio de um sistema de marcação chamado Barcode. Esta funcionalidade requer um processo computacional para separação dos dados por amostra, pois, o sequenciador fornece a mistura de todas as amostras em uma única saída. Este processo deve ser seguro a fim de evitar eventuais embaralhamentos que podem prejudicar as análises posteriores. Neste contexto, percebeu-se a necessidade de desenvolvimento de um modelo probabilístico capaz de atribuir um grau de confiança ao sistema de marcação utilizado em sequenciamentos multiplex. Os resultados obtidos corroboraram a suficiência do modelo obtido, o qual permite, dentre outras coisas, guiar um processo de filtragem dos dados e avaliação dos protocolos de sequenciamento utilizados.

## ABSTRACT

The next-generation sequencers such as Illumina and SOLiD platforms generate a large amount of data, commonly above 10 Gigabytes of text files. Particularly, the SOLiD platform allows the sequencing of multiple samples in a single run, called multiplex run, through a tagging system called Barcode. This feature requires a computational process for separation of the data sample because the sequencer provides a mixture of all samples in a single output. This process must be secure to avoid any harm that may scramble further analysis. In this context, realized the need to develop a probabilistic model capable of assigning a degree of confidence in the marking system used in multiplex sequencing. The results confirmed the adequacy of the model obtained, which allows, among other things, to guide a process of filtering the data and evaluation of the sequencing protocol used.

## 1. Introdução

O processo de sequenciamento genético surgiu em 1975, com a criação do método Sanger, desenvolvido por F. Sanger e, independentemente, por A. Maxam e W. Gilbert, como dito em Lesk (2008). Este método permite o sequenciamento de pequenos pedaços de DNA, não mais que 500 nucleotídeos, unidade básica de informação que constitui o DNA/RNA.

A combinação destes nucleotídeos guarda a informação gênica de um organismo, ou seja, a sequência de seu genoma, denominado Genótipo. O objetivo do processo de sequenciamento é determinar a ordem das bases nucleotídicas e confrontar com as características presentes no indivíduo como a predisposição ao desenvolvimento de câncer de mama, hipertensão arterial ou simplesmente características físicas como cor dos olhos e formato do crânio, assim exposto em Thompson & Thompson (2008).

Atualmente, existem sequenciadores da classe *High-Throughput Sequencing* também chamados de sequenciadores de nova geração, como as plataformas 454/Roche, Illumina/Solexa e ABI SOLiD. Estes sequenciadores produzem milhões de pequenas leituras de sequências (de 35 a 400 bases nucleotídicas) de diferentes amostras em apenas uma única corrida, denominada por Harismendy (2009) como corrida *multiplex*.

Particularmente, a plataforma ABI SOLiD produz milhões de pequenas leituras (35 bases em sua Versão 2) codificadas em "*2 base color enconding*", também chamada de *colorspace*, onde as transições entre duas bases nucleotídicas são representadas por uma cor característica obtida pela detecção dos fluorocromos: FAM, Cy3, TXR, ou Cy5; como exposto em Breu (2010). O que, via de regra, acaba por implicar na necessidade de se adicionar um passo para conversão dos dados, a fim de se obter a sequência do DNA nucleotídeos. A Figura 1 é adaptada de Breu (2010) e apresenta as cores utilizadas na conversão entre nucleotídeos e *colorspace* por meio da detecção dos fluorocromos.

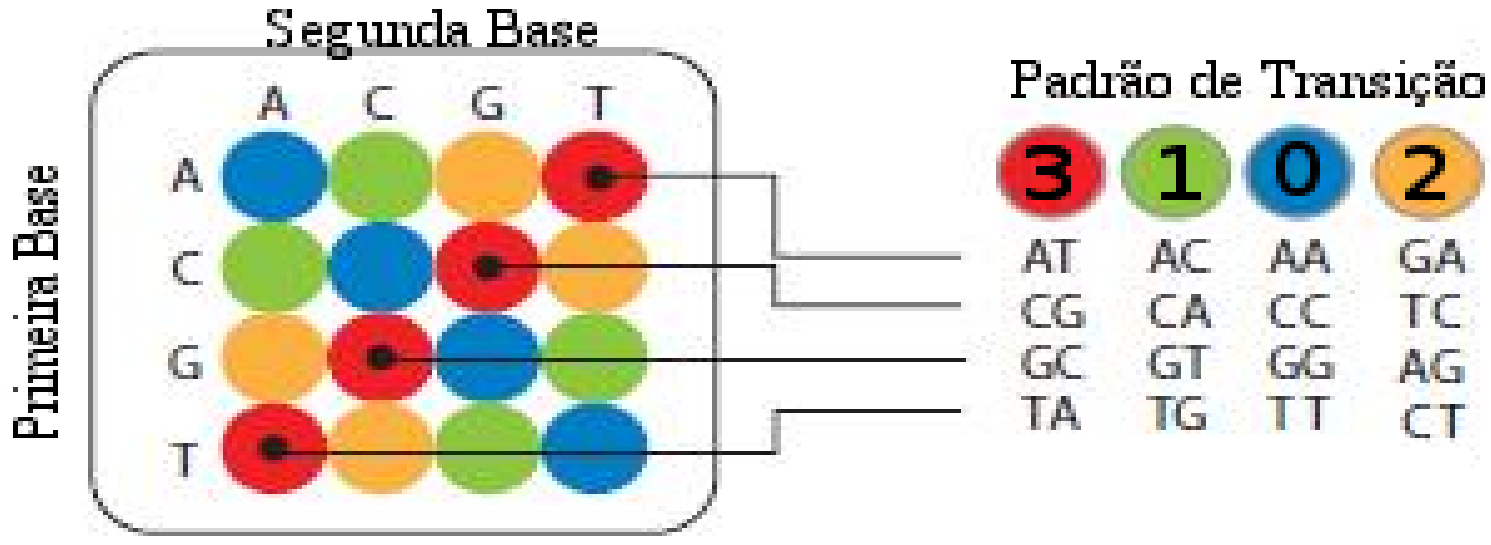

**Figura 1: Esquema visual de conversão entre nucleotídeos e *colorspace.***

No processo de conversão para *colorspace*, mantém-se a primeira base nucleotídica da sequência e inicia-se o passo de conversão seguindo-se a lógica apresentada na Figura 1. Por exemplo, na conversão da sequência GGCAG para *colorspace*, mantém-se o primeiro nucleotídeo G e verifica o código referente à transição GG, que é 0, então adiciona-se à sequência convertida, G0, a próxima avaliação será realizada na dupla GC, consultando a Figura 1, sabe-se que o código referente é 3, portanto, G03. Esse processo é realizado até a última base, a sequência final em *colorspace*, G0312.

Adicionalmente ao sistema de codificação próprio, o SOLiD suporta sequenciamento *multiplex* de até 256 amostras, através, dentre outros itens, do sistema de marcação denominado pela Applied Biosystem (2010) de *barcodes*. É importante que este sistema de marcação seja confiável para que não ocorra embaralhamento das amostras, o que representaria em uma perca de recursos computacionais para processamento de dados incongruentes e até mesmo apresentação de resultados aos especialistas, no caso, biólogos e biomédicos.

Para evitar estas falhas, alguns trabalhos presentes na literatura atual como Sassom (2010) e Salmela (2010) foram desenvolvidos para filtrar e tratar erros dos dados provindos da Plataforma ABI SOLiD, respectivamente. Contudo, o sistema de marcação possui características intrínsecas e requerem um tratamento especial para caracterização, filtragem e tratamento dos dados.

Sendo assim, o diferencial do presente estudo é o desenvolvimento de um modelo matemático baseado em Cadeias de Markov e realização de uma análise estatística a fim de caracterizar o sequenciamento *multiplex* quanto ao sistema de marcação. Os resultados obtidos permitiram constatar que as a modelagem desenvolvida e as medidas-resumo selecionadas são suficientes para caracterizar os sequenciamentos e, consequentemente, guiar o processo de filtragem dos dados e de análise protocolar do sequenciamento.

Este artigo está organizado como segue: a seção dois descreve brevemente a plataforma de sequenciamento que produziu os dados utilizados neste trabalho, a seção três descreve o modelo probabilístico desenvolvido, a seção quatro apresenta os resultados obtidos enquanto a quinta e última seção aborda as considerações finais.

## 2. Materiais e Métodos

Nesta seção serão apresentadas algumas características intrínsecas da plataforma ABI SOLiD, noções sobre o processo de preparação de amostras para sequenciamento e, por fim, o estudo de caso utilizado nos experimentos.

### 2.1. A plataforma ABI SOLiD

Os materiais biológicos utilizados neste trabalho são provenientes de sequenciamentos *multiplex* da Plataforma SOLiD. Nas corridas *multiplex*, o sequênciador gera quatro arquivos: um corresponde ao sistema de marcação das amostras, contendo a sequência de marcação; outro corresponde a sequência das amostras, também chamada de sequência de interesse; e dois com a qualidade associada às sequências de marcação e de interesse.

Os arquivos que representam as sequências obtidas estão no formato “csfasta”; e os que contêm o valor da qualidade associada a cada transição de base das sequências lidas estão no formato “qual”. A fim de realizar análises biológicas posteriores, é necessário separar as sequências de interesse por amostra. A tabela I apresenta a sintaxe padrão dos arquivos e um exemplo contendo uma sequência de marcação.

**Tabela 1. Sintaxe padrão dos arquivos de saída da plataforma ABI SOLiD.**

| Sintaxe | Exemplo |
|---|---|
| #Cabeçalho<br>>Identificador_R3<br>Sequência de marcação ou de Interesse ou Qualidade | # Title: miRNA_20090819_1<br>>1_44_108_R3<br>G01130 |

O identificador (ID) é o mecanismo que permite, na etapa de separação por amostra, ligar as informações do sistema de marcação às informações da sequência de interesse. No identificador, o sufixo, que pode ser “_R3” ou “_F3” indica que a informação da linha posterior refere-se ao sistema de marcação ou à sequência de interesse; o prefixo é que, de fato, é o identificador que permitirá a busca da sequência de interesse quando aquele ID for classificado como de uma determinada amostra.

Na coluna Exemplo da Tabela 1, temos o identificador “>1_44_108_R3”, pelo sufixo sabe-se que a informação da linha posterior refere-se ao sistema de marcação, e a sequência de marcação “G01130” possui *match* com o *Barcode 6* da biblioteca padrão, como exposto em Ambion (2009). Portanto, para separar as informações das sequências de interesse, um *script* deverá buscar o ID “>1_44_108” nos arquivos F3 e indexar as informações em um arquivo próprio da amostra marcada com o *Barcode 6*. A plataforma ABI SOLiD possui, acoplado ao sequênciador, um c*luster* para pré-processamento dos dados em relação à qualidade, contudo, não há como fazer o pareamento dos arquivos em relação ao ID, sendo necessária esta etapa no processamento posterior dos dados.

## 2.2. O processo de sequenciamento

Para um melhor entendimento do estudo genômico de um determinado organismo é necessário conhecer o processo de sequenciamento. Este processo é antecedido de alguns procedimentos essenciais onde após a coleta das amostras, o DNA deve ser fragmentado, pois o sequenciador pode ler somente fragmentos de 35 a 50 pares de base; aos fragmentos, são adicionados dois adaptadores, um responsável por aderir o fragmento a uma estrutura metálica denominada *bead*, e o segundo adaptador adere o fragmento a um material flutuante para que ele seja retirado das lâminas de sequenciamento, caso não ocorra a aderência usando o primeiro adaptador, evitando assim possíveis contaminações. A Figura 2 apresenta um esboço geral da preparação das amostras para a plataforma SOLiD.

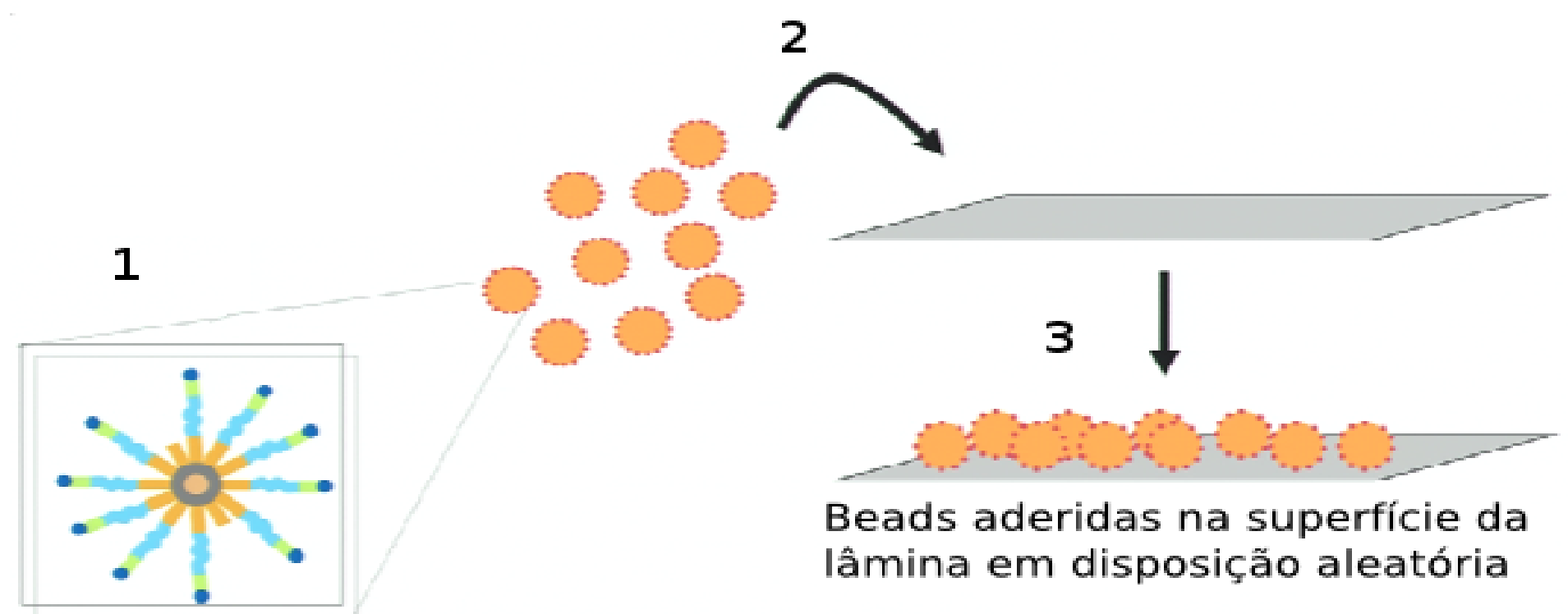

**Figura 2: Esboço da preparação de amostras para submissão à plataforma SOLiD.**

Na Figura 2 observa-se em 1 uma *bead* metálica com várias cópias de um determinado fragmento da amostra. Isto é obtido por meio da amplificação do fragmento por uma técnica conhecida por *Polymerase Chain Reaction* (PCR). Isto é necessário, pois, os sequenciadores não conseguem obter leitura da sequência de um único fragmento. A mesma figura apresenta no passo 2, o depósito das *beads* contendo os fragmentos obtidos da amostras. Em seguida, no passo 3, observa-se a lâmina pronta para submissão ao SOLiD.

## 2.3. Estudo de caso

O material biológico utilizado nos estudos para caracterização do sistema de marcação na plataforma SOLiD foi proveniente de pacientes com câncer. Todos forneceram consentimento por escrito e o estudo foi aprovado pelo Comitê de Ética em pesquisa (CEP) do Hospital Universitário João Barros Barreto (HUJBB) – Universidade Federal do Pará (UFPa), protocolo número 14052004 / HUJBB. Utilizaram-se cinco pacientes, dos quais foram extraídas duas amostras cada.

Para a preparação dos fragmentos das amostras, utilizou-se o SOLiD Small RNA Expression Kit, o qual encontra-se descrito em Ambion (2009). Todos os miRNAs da biblioteca foram anexados com uma curta sequência de DNA usada para definir o trecho do DNA a ser amplificado, no caso, o sistema de *barcodes*. Neste contexto, foram realizados dois sequenciamentos, onde gerou-se 300 Gigabytes (Gb) de dados, dos quais cerca de 50 Gb correspondiam aos arquivos "csfasta" e "qual", descritos na subseção 2.1 deste artigo.

## 3. Modelagem Probabilística

Mediante análise preliminar dos dados disponíveis e de pesquisa na literatura atual, verificou-se uma lacuna no desenvolvimento de uma modelagem probabilística para o problema de avaliação e caracterização de corridas *multiplex* na plataforma SOLiD, onde a maioria dos trabalhos utilizam-se de heurísticas para filtragem e recuperação dos dados desta plataforma.

Desta forma, percebeu-se a necessidade do desenvolvimento de uma modelagem probabilística para o problema, assim como, a busca de uma estatística suficiente para a caracterização do sequenciamento com relação ao sistema de marcação. O que possibilita o desenvolvimento de um filtro customizado para os dados oriundos de corridas *multiplex,* a avaliação dos protocolos de sequenciamento utilizados para preparação das amostras.

A modelagem probabilística baseou-se no valor de qualidade que está contido nos dados do sequenciador, este valor está compreendido no intervalo de -1 à 35. Por motivos mercadológicos, a Applied Biosystem não disponibiliza como este valor é obtido ou a função de mapeamento entre o valor da qualidade e a probabilidade associada. Contudo, em (Ewing, 1998) encontrou-se uma função que, adaptada ao intervalo dos valores de qualidade informados pela plataforma SOLiD, consegue aproximar o mapeamento qualidade-probabilidade, como segue:

$$P(Q) = 1 - 10\exp((-(Q+1)/10)) \qquad (1)$$

Como o intervalo do valor de qualidade inclui um número negativo, necessitou-se transformar este intervalo por meio da adição de uma unidade ao valor de qualidade (Q+1). O valor obtido em P(Q) representa a probabilidade associada à leitura da transição de duas bases, a qual gera um estado em *colorspace*. Para avaliar o grau de confiança de uma sequência, modelou-se a sequência de marcação como uma Cadeia de Markov. A base inicial da sequência e as transições posteriores representam estados, e a probabilidade de transição de estados é o resultado obtido em (1).

Para elucidar melhor esta abordagem, tomemos o exemplo da seguinte sequência dada em *colorspace*: G00, com valor de qualidade "10 24". Por meio do mapeamento qualidade-probabilidade proposto em (1), a probabilidade das transições seriam: 0,92057 e 0,99684, como apresentado na Figura 3.

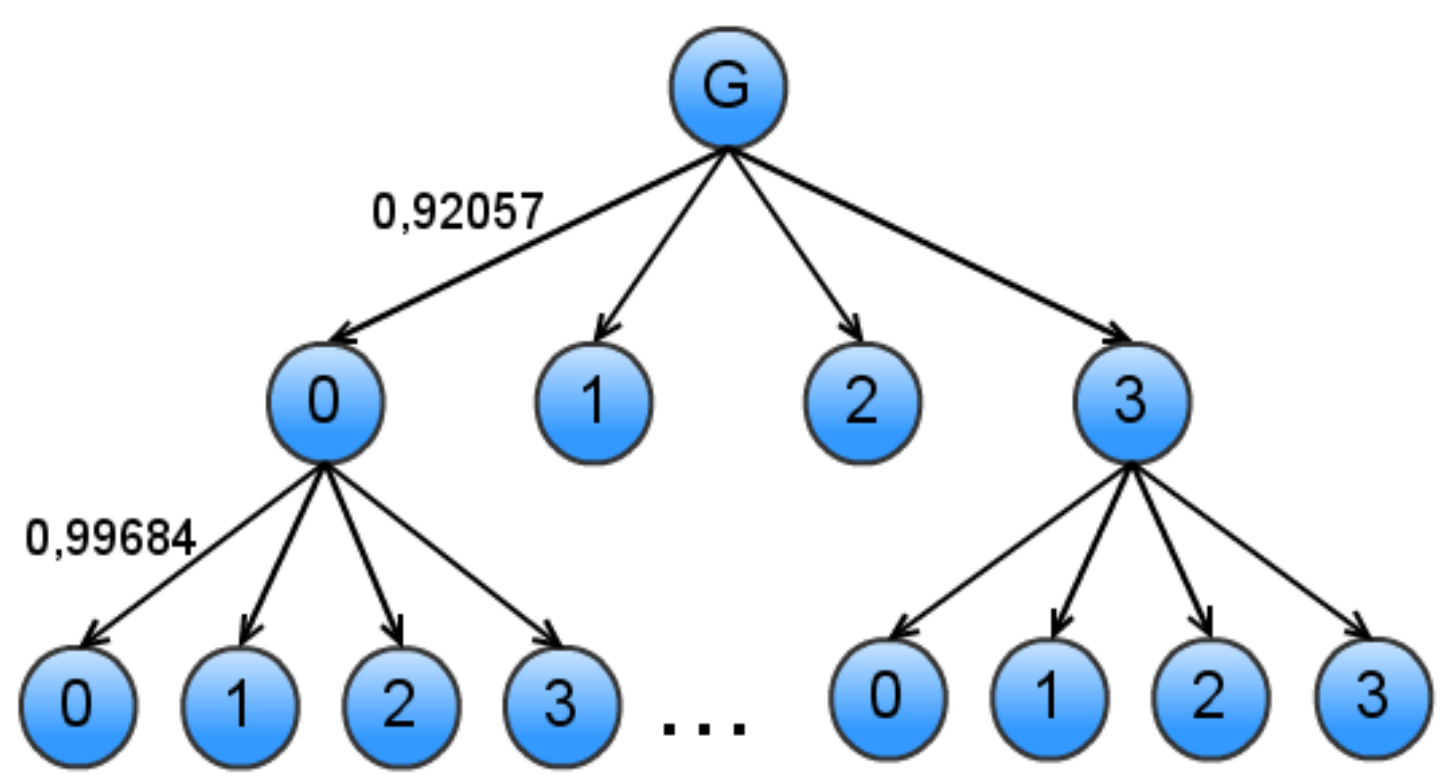

**Figura 3: Grafo da CM que representa a sequência "G00" com qualidade associada "10 24".**

Para calcular o grau de confiança de uma determinada sequência θ, ou seja, a probabilidade desta sequência não ter sido produto do acaso, multiplicam-se as probabilidades de todas as transições existentes, como apresentado em (2). O resultado representa a probabilidade de a sequência obtida ser, de fato, a presente na amostra; no exemplo da figura 3, a probabilidade da sequência G00 não ser produto do acaso é de 0,91766.

$$P(\theta) = \prod P(i) \qquad (2)$$

Este modelo probabilístico permite, dentre outros pontos, a descoberta de medidas-resumo capazes de caracterizar o sequenciamento quanto ao sistema de marcação; guiar um processo de filtragem adaptativa, respeitando as características intrínsecas de cada corrida *multiplex;* e desenvolver um processo de recuperação de sequências que não correspondem exatamente com algum *barcode* da biblioteca padrão, baseando-se em métodos de Inferência Bayesiana. As duas primeiras aplicações, descoberta de medidas-resumo e filtragem adaptativa são abordadas nas seções posteriores.

## 4. Apresentação dos Resultados

A busca por medidas resumo de uma determinada série de amostras é uma tarefa não-trivial, tendo em vista que os valores encontrados devem ser suficientes e representativos de toda a amostra, como apresentado em Bussab e Morettin (2002). No contexto da análise de dados oriundos da plataforma SOLiD, somente medidas de posição, como média e moda, e medidas de dispersão, como desvio padrão e variância, não são suficientes para descrever uma corrida *multiplex,* pois, informações como taxa de sequências de marcação que obtiveram *match* com os *barcodes* da biblioteca padrão e número de sequências não pareadas entre os arquivos; são importantes para a análise da qualidade de uma corrida multiplex. Esta última medida se dá ao fato de os arquivos serem divididos, como apresentado anteriormente, em dois arquivos com informações sobre o sistema de marcação e dois arquivos com informações sobre as sequências de interesse.

Para encontrar as medidas-resumo capazes de caracterizar o sequenciamento *multiplex,* testes computacionais foram realizados e os resultados obtidos foram avaliados por um especialista no domínio e considerados consistentes. Por meio deste processo, definiram-se as seguintes medidas resumo para caracterização do sequenciamento: número de sequências de marcação lidas; número de sequências de interesse lidas; número de identificadores não pareados entre marcação e sequências de interesse; média, moda, variância, taxa sequências com *match* com os *Barcodes* e taxa de intersecção dos identificadores. Para melhor visualizar as medidas estatísticas geraram-se gráficos das funções de densidade de probabilidade de cada corrida, como apresentado na figura 4.

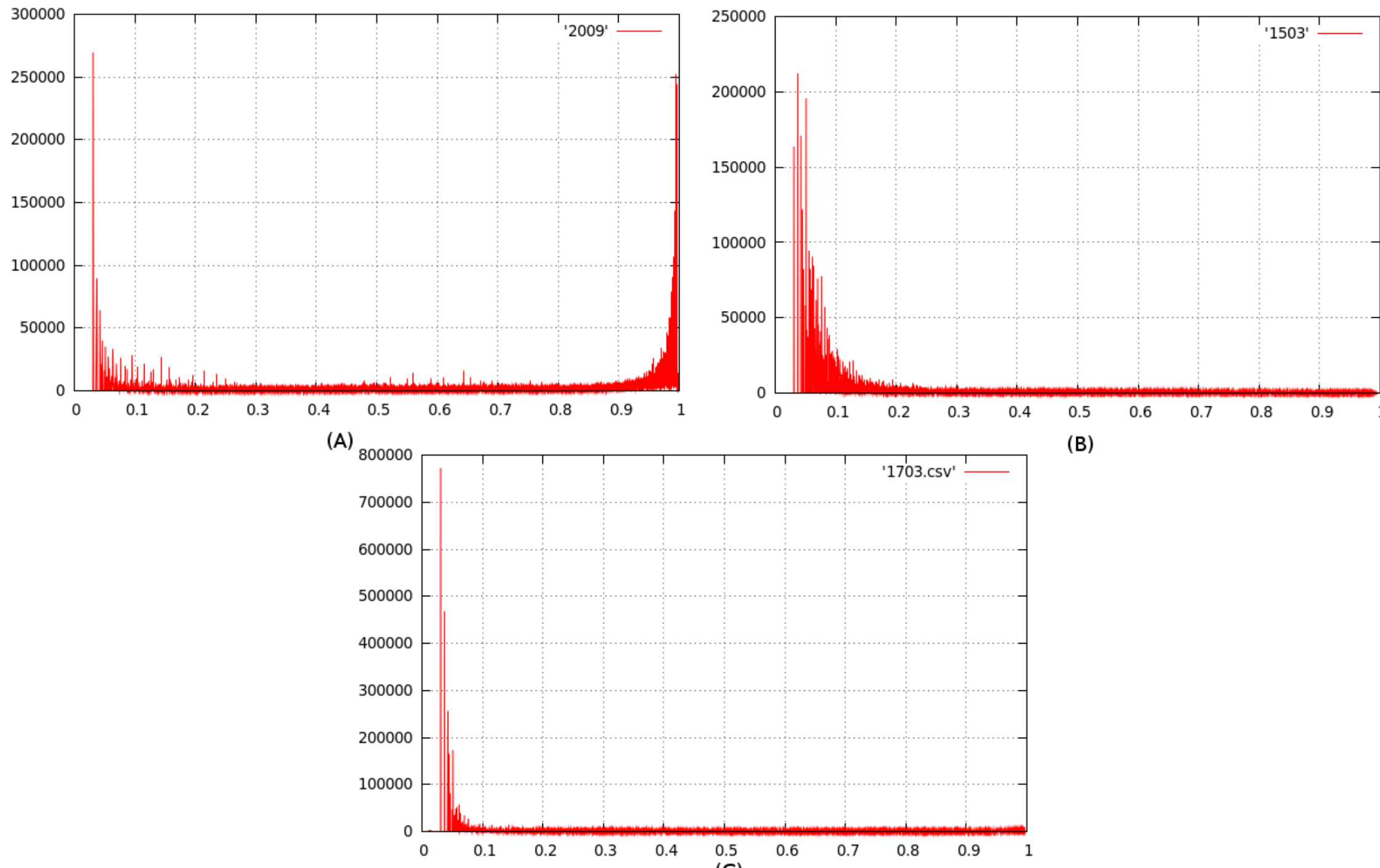

**Figura 4: (A) Gráfico correspondente à Corrida 1; (B) Gráfico correspondente à Corrida 2; (C) Gráfico correspondente à Corrida 3.**

Por meio da análise gráfica da Figura 4(A), percebe-se uma alta concentração de leituras com boa qualidade próximas de 1; no entanto, a Figura 4(B) apresenta alta densidade próximas a probabilidade 0, o que implica em uma corrida com baixa confiança. Vale ressaltar que, apesar da figura 4(C) aparentemente apresentar leituras de baixa qualidade, escala do eixo x é maior que as outras escalas, pois, a frequência da moda neste gráfico é consideravelmente superior que nos outros gráficos. A fim de facilitar a interpretação dos resultados, a Tabela 2 descreve as medidas-resumo obtidas para as três corridas.

**Tabela 2: Informações sobre os sequenciamentos.**

| | Corrida 1 (C1) | Corrida 2 (C2) | Corrida 3 (C3) |
|---|---|---|---|
| Número de leituras das sequências de marcação | 142.453.565 | 29.602.637 | 51.832.867 |
| Número de leituras das sequências de interesse | 158.673.424 | 19.523.612 | 27.634.981 |
| Número de IDs não pareados entre marcação e sequências de interesse | 23.631.427 | 7.593.087 | 13.147.217 |
| Média | 0,8123 | 0,4715 | 0,5917 |
| Variância | 0,0553 | 0,0687 | 0,0806 |
| Moda | 0,3088 | 0,3726 | 0,3088 |
| Taxa de sequências de marcação com *match* com *barcodes* | 81,44% | 1,63% | 48,18% |

As três primeiras informações na tabela representam o número de sequências de marcação, sequências de interesse; e o número de identificadores não pareados entre marcação e sequência de interesses. As duas primeiras deveriam ser idênticas, pois, as sequências de marcação permitem à separação das sequências de interesse por amostra, logo, elas estão associadas, todavia, falhas na plataforma fazem com que estes números apresentem uma discrepância.

Este fato implica na necessidade de se parear os arquivos contendo as informações do sistema de marcação com os arquivos contendo as informações da sequência de interesse. A diferença entre os identificadores pareados e não pareados foi selecionado como o terceiro atributo da Tabela 2, o número de identificadores não pareados entre marcação e sequências de interesse.

Outras informações como média; moda e variância; auxiliam, dentre outros pontos, na análise do protocolo de preparação de amostras para corrida *multiplex,* na elaboração e adoção de uma estratégia de filtragem dos dados. Por exemplo, em C1 dados como média (0,8122), variância (0,0553) e taxa de sequências de marcação com *match* com *barcodes* (81,44%) permitem afirmar que esta corrida apresentou boa qualidade quanto ao sistema de marcação, mesmo a moda ser um valor que representa baixa qualidade dos dados, 0,3088.

Na Corrida 2, obteve-se o pior resultado, primeiro com uma baixa taxa de sequências de marcação com *match* com *barcodes,* apenas 1,63%; depois com probabilidade de certeza inferior à 50%, o que indica uma alta probabilidade de, a sequência obtida, ter sido ao acaso. Em resultados similares indica-se a revisão dos protocolos utilizados, testes de qualidade dos reagentes utilizados e não utilização dos dados para as análises biológicas posteriores.

Por fim, o terceiro sequenciamento apresenta qualidade razoável, com grau de confiança de 0,5917 e variância considerável neste domínio de aplicação – 0,0806, logo, há leituras com um bom grau de confiança; para plena utilização destes dados indica-se um processo de filtragem criterioso e reavaliação do resultado destes processos em relação à Taxa de sequências de marcação com *match* com *barcodes.*

A combinação destas informações permite, dentre outras coisas, avaliar o desempenho do *cluster* acoplado ao sequênciador, que é responsável por pré-filtrar, parear e gravar os dados gerados; avaliar o protocolo de sequenciamento utilizado para preparação das amostras, no tocante à adição da sequência de marcação; guiar o desenvolvimento de um filtro customizado para corridas *multiplex*, reduzindo o custo computacional das etapas posteriores; e por fim, aumentar a confiabilidade do conhecimento gerado.

## 5. Considerações finais

A falta na literatura de medidas-resumo capazes de caracterizar o sequenciamento quanto ao sistema de marcação motivou este estudo. Ressalta-se a importância de um alto grau de confiabilidade destes dados em particular, pois, falhas no sistema de marcação podem ocasionar o embaralhamento das sequências de interesse, o que implica em desperdício de poder computacional para as análises genômicas bem como eventuais erros nos resultados obtidos.

Dentre as contribuições deste estudo, a principal o desenvolvimento de um modelo baseado em uma Cadeia de Markov capaz de obter o grau de confiança de uma sequência provinda do sequenciador SOLiD. Análises estatísticas posteriores permitiram também a identificação de medidas-resumo capazes de caracterizar o perfil da qualidade dos barcodes obtidos em uma corrida; são elas: média; moda; variância; comparação entre o número de leituras obtidas do sistema de marcação e das sequências de interesse; o número de identificadores pareados entre estes dois dados; e taxa de sequências de marcação com *match* com *barcodes*.

Tais informações possibilitam a análise do protocolo de bancada utilizada no sequenciamento; a avaliação de um possível descarte dos dados gerados, além disso, em um trabalho futuro, esta análise pode guiar o processo de desenvolvimento de um filtro customizado para os barcodes.

## Referências